\documentclass[a4paper]{amsart}

\usepackage{amsmath} 
\usepackage{amsfonts} 
\usepackage{latexsym} 
\usepackage{graphicx} 
\usepackage{multicol} 
\usepackage[all]{xy} 
 
\usepackage{mathrsfs} 
 
\newcommand{\absolute}[1]{\lvert#1\rvert} 
\newcommand{\mathset}[1]{{\left\{#1\right\}}}  
\DeclareMathOperator{\PGL}{PGL} 
 
\DeclareMathOperator{\Ends}{Ends}
\newcommand{\Const}{{\rm const}} 
\newcommand{\Even}{{\rm even}}
\newcommand{\Odd}{{\rm odd}}
\newcommand{\Id}{{\rm id}} 
\newcommand{\Child}{{\rm ch}}
\newcommand{\Image}{{\rm im}}
\newcommand{\Vol}{{\rm Vol}}
\newcommand{\Finite}{{\rm fin}} 
\newcommand{\Identity}{{\rm id}}
\newtheorem{Rem}{Remark}[section] 
\newtheorem{Exa}[Rem]{Example} 
\newtheorem{Def}[Rem]{Definition} 
\newtheorem{thm}[Rem]{Theorem}

\begin{document} 
 
\title{Mumford dendrograms} 
\author{Patrick Erik Bradley} 
\date{\today} 
 
 

\begin{abstract} 
An effective $p$-adic encoding of dendrograms is presented through an explicit
embedding into the Bruhat-Tits tree for a $p$-adic number field.
This field depends on the number of children of a vertex and is a finite
extension of the field of $p$-adic numbers. It is shown that fixing
$p$-adic representatives of the residue field allows a natural way
of encoding strings by identifying a given alphabet with such representatives.
A simple $p$-adic hierarchic classification algorithm is derived
for $p$-adic numbers, and is applied to strings over finite
alphabets. Examples of DNA coding are presented and discussed.
Finally, new geometric and combinatorial invariants of time series
of $p$-adic dendrograms
are developped.  
\end{abstract}

\maketitle 
 
%
%
 
\section{Introduction} 

A dendrogram is often the output of a hierarchical classification algorithm.
In the usual agglomerative methods, it is obtained from data by a distance
function which is adjusted after each iteration to the clusters obtained in the previous step. Classically, the distance is euclidean, and the hierarchical
 structure is fitted to the data. The analyst then has to decide by other means whether the resulting dendorgram represents the underlying hierarchical structure of the data, or not. In the $p$-adic world, however, there is no ambiguity concerning the interpretation of dendrograms. The reason is that the $p$-adic
distance is ultrametric. This has the effect that a $p$-adic dendrogram correctly represents the hierarchies within a given set of $p$-adic numebrs, of course
with respect to the $p$-adic metric. Another effect is, as we will show, that
$p$-adic classification is algorithmically much simpler than its classical counterpart. The consequence for data mining lies in the shift from classification to data encoding.

If the dendrogram $X$ is known, then its $p$-adic encoding can be effected by associating paths from the top cluster down to the data with $p$-adic numebrs. This is in fact an embedding of $X$ into the $p$-adic Bruhat-Tits tree which can 
be seen as a ``universal dendorgram''. This embedding will be made precise
in this article. Strings over an alphabet are the only instance known to the author, in which $p$-adic data encoding can be realised in a straightforward manner. The encoding depends on the coefficients in $p$-adic expansions associated to the alphabet. Examples of $p$-adic DNA encoding are proposed and discussed. 

Time series of $p$-adic dendrograms give rise to  new geometric invariants.
namely, if translations along geodesic lines in the Bruhat-Tits tree can be identified, a discrete group action can be estimated in important cases.
This action then leads to a dynamic system on a so-called {\em Mumford curve},
the $p$-adic analogon of a riemann surface. Studying this dynamic system will 
yield
pararmeters whic can be used e.g.\ for extrapolating hierarchical data in time.

Possible applications of $p$-adic dendrograms are coding theory of graphs and
strings. Another area of application can be spatial reasoning and querying, including space-time issues. The time series point of view is naturally applicable to strings.
The idea of studying $p$-adic dendrograms is taken from \cite{MurtaghJoC2004}.
Linear fractions are considered in \cite{DKM2006} in the $p$-adic and real case
simultaneously. A description for a general audience of the $p$-adic Bruhat-Tits tree and some of its
discrete symmetries can be found in \cite{CK2005}.

 
\section{Embedding a dendrogram into the $p$-adic Bruhat-Tits tree} 
 
In order to embed a dendrogram $X$ into the $p$-adic Bruhat-Tits tree, 
we first define $X$ as the dendrogram for its data plus an extra point
$\infty$. The reason is that in this way the top cluster becomes
the vertex uniquely determined by $\infty$  and two data points at maximal
distance. This viewpoint leads to the term {\em projective dendrogram},
and we will see that in the $p$-adic case, it is associated to the $p$-adic
projective line minus the $p$-adic numbers representing the data and $\infty$.

\subsection{Abstract dendrograms} \label{abstractdendro}
 
Dendrograms represent hierarchies within data, and are therefore trees, 
i.e.\ graphs without loops.  
Subsets of data points are clusters represented by the vertices, and  
inclusions of clusters are represented by paths between the corresponding vertices. 
It is useful to distinguish between clusters and data in the same way 
as one distinguishes sets from their elements: even if certain 
clusters are singletons, they are nevertheless not data in the same way 
as the set $\mathset{x}$ is in a strict sense not the same 
thing  as the point $x$. Hence, in our viewpoint, the data will not be 
part of, but at the boundary of a dendrogram. Hence, we will allow 
graphs to have unbounded edges. 
 
\begin{Def} 
A {\em graph} is a quadruple $\Gamma=(\Gamma^0,\Gamma',\partial,\iota)$, 
where $\Gamma^0$ and $\Gamma'$ are sets, $\partial\colon\Gamma'\to\Gamma^0$ 
is a map, and $\iota\colon\Gamma'\to\Gamma'$ is an idempotent map 
(i.e.\ $\iota\circ\iota=\Id$). 
The elements of $\Gamma^0$ are called {\em vertices}, and those of $\Gamma'$
{\em flags}. $\partial$ is called the {\em boundary map}, and  $\iota$ the
{\em inversion}. A graph $\Gamma$ is {\em finite} if   $\Gamma^0$ and $\Gamma'$ are both finite.  
\end{Def} 
 
The inversion yields an equivalence relation on the set of flags: 
$F_1\sim F_2$ iff $F_1=F_2$ or $F_1=j(F_2)$. The equivalence classes 
under $\sim$ are called the {\em edges} of $\Gamma$. The set of edges is 
denoted by $\Gamma^1=\Gamma'/\sim$. An edge  
is called {\em unbounded} if it consists of a single flag, otherwise it is
called {\em internal}. We denote the set of internal resp.\ unbounded 
edges by $\Gamma^1_0$ resp.\ $\Gamma^1_\infty$.

A graph $\Gamma$ has a {\em topological model} $\absolute{\Gamma}$, 
obtained by identifying each flag $F$ with the half-open interval $[0,1)_F$ 
and pasting $\partial F$ to $F$, and then taking the quotient by $\sim$.
This model reflects important topological properties of the graph,
such as the number of connected components or the number of ``holes'', i.e.\
minimal loops in $\Gamma$. These quantities are known as the {\em Betti 
numbers} $h_0(\absolute{\Gamma},\mathbb{R})$ and 
$h_1(\absolute{\Gamma},\mathbb{R})$ from algebraic topology,
where they are introduced as the dimension of certain real vector spaces. 
For
finite graphs, there is an important formula which relates the Betti numbers to
the combinatorial data:
$$
h_0(\absolute{\Gamma},\mathbb{R})-h_1(\absolute{\Gamma},\mathbb{R})
=\#\Gamma^0-\#\Gamma^1_0,
$$
known as the {\em Euler formula}.

A graph $\Gamma$ is a {\em tree} if it is connected and
without loops, or, equivalently, if the Betti numbers of $\absolute{\Gamma}$
satisfy
 $h_0(\absolute{\Gamma},\mathbb{R})=1$ and 
$h_1(\absolute{\Gamma},\mathbb{R})=0$.
 
A {\em rooted tree} is a pair $(T,v)$, where $T$ is a tree and $v\in T^0$ is a
vertex. The distinguished vertex $v$ is called the {\em root} and
 makes a rooted tree $(T,v)$ into a
directed
tree by orienting all edges away from $v$, i.e.\ an internal edge $e$ with
boundary $\mathset{w_1,w_2}$ is oriented from $w_1$ to $w_2$, if $w_1$
is closer to $v$ than $w_2$ is, and an unbounded edge $e'$ is oriented away
from the unique vertex $\partial{e'}$.

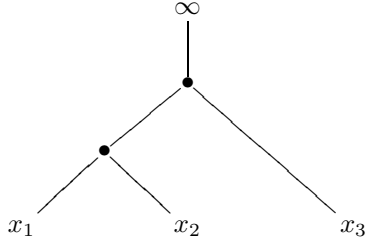
\begin{figure} 
$$ 
\xymatrix@=20pt{ 
&&\infty\ar@{-}[d]&&\\ 
&&*\txt{$\bullet$}\ar@{-}[dl]\ar@{-}[ddrr]&&\\ 
&*\txt{$\bullet$}\ar@{-}[dl]\ar@{-}[dr]&&&&\\ 
x_1&&x_2&&x_3 
} 
$$ 
\caption{Projective dendrogram with three datapoints.} \label{projdendro} 
\end{figure}

By an {\em abstract dendrogram} we mean a finite
rooted tree $\mathcal{T}=(T,v)$, all of whose 
vertices originate in at least two edges (unbounded or not). 
It is {\em labelled},
if its unbounded edges are labelled by some bijective map 
$\lambda\colon T^1_\infty\to L$, where $L$ is a set 
whose elements are called {\em labels}. A {\em projective dendrogram}
is a labelled abstract dendrogram $\mathcal{T}$ whose root originates in an
umbounded edge labelled $\infty$ and at least two more edges. The
unbounded edges not labelled $\infty$ are called the {\em datapoints} 
or {\em data} underlying $\mathcal{T}$, and will be denoted by $D$. 
Figure \ref{projdendro} illustrates a projective dendrogram with three data
points. The reason for calling it ``projective'' will become apparent in later
subsections.
 
\begin{Rem}
To call unbounded edges of a dendrogram $\mathcal{T}$ 
``data'' seems to be in contradiction
to the initial purpose of having data  not be part of a dendrogram.
However, by looking at the topological model, we see first 
that an unbounded edge
is nothing but a half-line in the tree $T$ underlying $\mathcal{T}$. 
The {\em ends} of $T$ are equivalence classes of halflines, where
halflines differring only in finitely many internal edges are equivalent.
According to graph theory, the ends 
form the boundary $\partial T$ of $T$. Hence, 
 we have in fact identified $\partial T\cong T^1_\infty$ 
with data and $\infty$.  
\end{Rem}
 
Given some projective dendrogram $\mathcal{T}=(T,v,\lambda,D)$, there is an order
relation $<$ on $T^1\setminus\lambda^{-1}(\infty)$: $e<e'$, if $e$ lies
on the path from $\infty$ to $e'$. If $e$ and $e'$ originate in some vertex
$v$, impose any total order $<_v$ on the edges originating in $v$ in
order to break ties. This extends to a total order on data as follows:
the lexicographic order with respect to $<$ and all $<_v$ on the reduced words
$\infty\cdots e$ associated to minimal paths $w_e$
(i.e.\ paths without backtracking)  from $\infty$ to $e$
induces a total order on $D$ via the unique bijection between $D$
and $\mathset{w_e\mid e\in T^1_\infty}$ induced by $\lambda$.
A projective dendrogram together with a total ordering $<$ 
on its data is called
{\em ordered}. We call a minimal path in a tree {\em geodesic}.

A {\em metric} on a projective dendrogram is a function $\mu\colon
T^1_0\to\mathbb{N}\setminus\mathset{0}$, and defines in an obvious manner
a distance $d\colon T^0\times T^0\to\mathbb{N}$. 
This induces a {\em level structure}
$\ell\colon T^0\to\mathbb{N},\;w\mapsto\ell(w)=d(v,w)$,
where $v$ is the root. Figure \ref{Murtaghdendro} displays
a projective dendrogram with level structure and $\infty$ on top.

\subsection{The binary case} \label{binary}

Let $X=(T,v,D,<,\mu)$ be an ordered projective dendrogram with metric.
In this subsection, we assume that $T$ is a binary tree, i.e.\
each vertex $w\in T^0$ has precisely two (internal or unbounded)  
directed outoing edges $e_0(w)<e_1(w)$. With $e_0:=e_0(v)$ and $e_1:=e_1(v)$
we have that $T\setminus\mathset{v}$ is the disjoint union of the two branches
$\Gamma_0\ni e_0$ and $\Gamma_1\ni e_1$. $\Gamma_0$ and
$\Gamma_1$ are themselves 
projective dendrograms, if the $e_i$ are labelled $\infty_i$.
We define functions
\begin{align*}
\chi_0\colon\Gamma_0^1\to\mathset{0,1},\;
&e\mapsto\begin{cases}
  0,&\exists w\in T^0\colon e=e_0(w)\\
  1,&\text{otherwise}
\end{cases}\\
\chi_1\colon\Gamma_1^1\to\mathset{0,1},\;
&e\mapsto\begin{cases}
  0,&\exists w\in \Gamma_1^0\colon e=e_1(v)\\
  1,&\text{otherwise}
\end{cases}
\end{align*}
Together, $\chi_0$ and $\chi_1$ define a function $\chi\colon T^1\to\mathset{0,1}$ such that $\chi(e_0)=0$ and $\chi(e_1)=1$. This extends to a function
on the set $\mathcal{G}(\infty,D)$ of directed geodesics $\gamma_x$
from $\infty$ to any datum $x\in D$
$$
\chi\colon\mathcal{G}(\infty,D)\to\mathbb{Q}_2,\;
\gamma_x\mapsto\sum\limits_{e\in\gamma_x}\chi(e)2^{\ell(o(e))},
$$
where $o(e)$ denotes the origin vertex of the edge $e$, and $\ell$
is the level function on $X$.
Together with the identification $\mathcal{G}(\infty,D)\cong D$, 
we obtain the $2$-adic encoding $\chi\colon D\to\mathbb{Q}_2$ of binary data. 

\begin{Rem}
The coding function $\chi$ is in fact $\mathbb{Z}_2$-valued. 
Even more, its values are natural numbers, because $X$
is finite. By construction, the values $0$ and $1$ are taken by $\chi$ for any projective
dendrogram. In fact, $\chi(x_0)=0$ and $\chi(x_n)=1$, if 
$D=\mathset{x_0<\dots<x_n}$.
\end{Rem}

\begin{Exa}
\begin{figure}
\centerline{
\includegraphics[scale=.28]{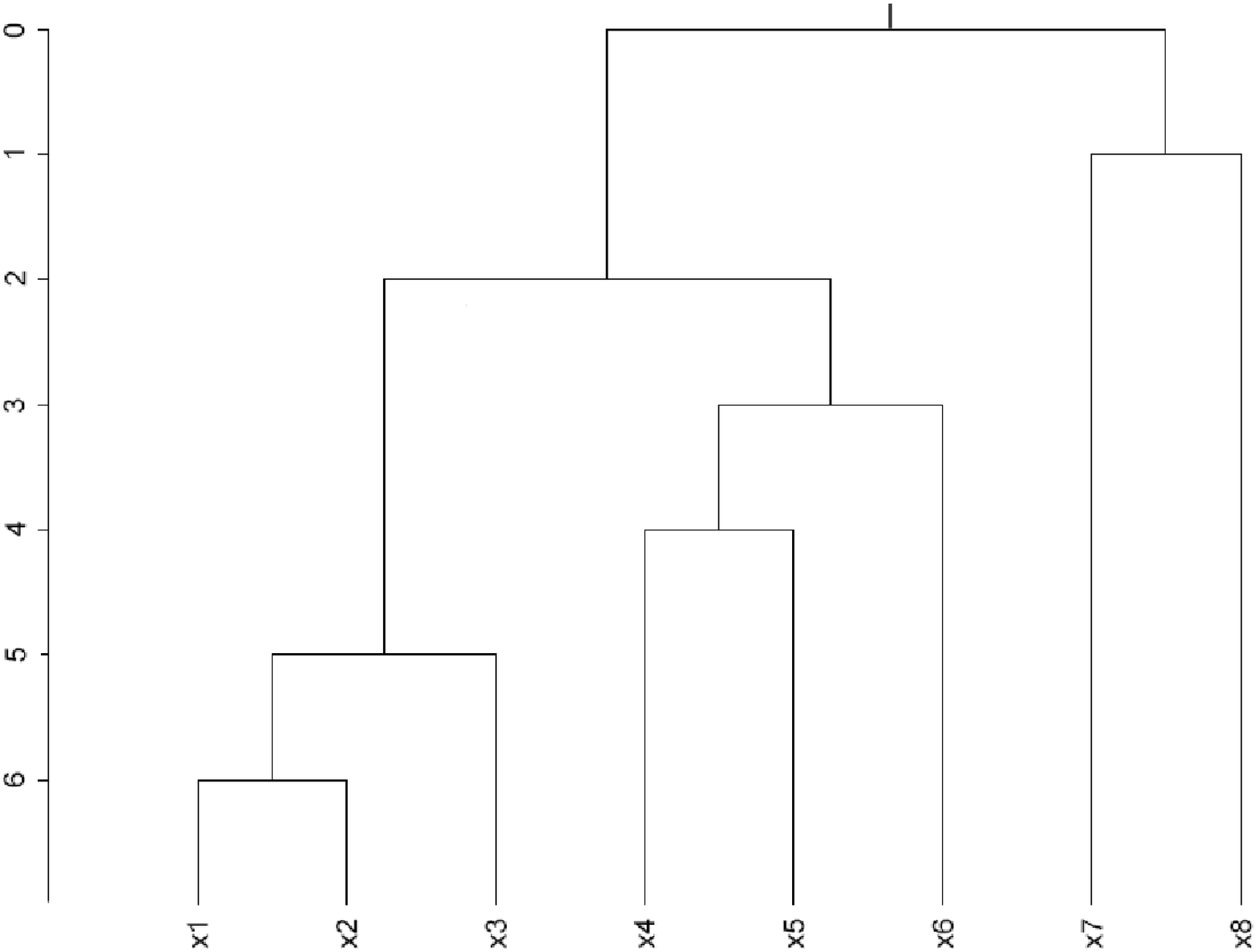}
}
\caption{A projective dendrogram with level structure.}
\label{Murtaghdendro}
\end{figure}

The dendrogram in Figure \ref{Murtaghdendro} has the $2$-adic encoding
$\chi$ given by:
$$
\begin{array}{llll}
x_1=0,&  x_2=2^6,&  x_3=2^5,& x_4=2^2\\
x_5=2^2+2^4,& x_6=2^2+2^3,& x_7=2^0+2^1,& x_8=1.
\end{array}
$$
This encoding differs slightly from the $2$-adic encoding of
the same dendrogram in \cite{Murtagh2004}.
\end{Exa}

\subsection{The Bruhat-Tits tree for $p$-adic fields} \label{btt}

It was observed that
an encoding of dendrograms with $p$-adic numbers from 
$\mathbb{Q}_p$ leads to
considering subtrees of the Bruhat-Tits tree
$\mathscr{T}_{\mathbb{Q}_p}$ \cite{BradDegendendrofam, BradDendrofam}.
Here, we intend to prepare an effective  embedding of dendrograms
into the Bruhat-Tits tree, which
 is going to be made  precise in Section \ref{p-ad-dendro}.
The preparation consists in  reviewing 
the construction of $\mathscr{T}_{\mathbb{Q}_p}$ and the variants for
finite extension fields $K$ of $\mathbb{Q}_p$, as the latter turns out too
small in general for encoding data. 

\subsubsection{The Bruhat-Tits tree for $\mathbb{Q}_p$}

\begin{figure}
\centerline{
\includegraphics[scale=.3]{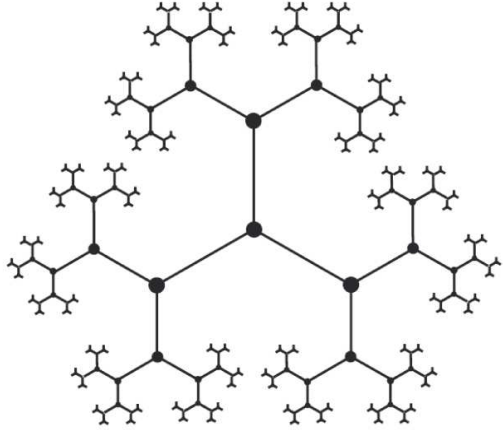}
}
\caption{The Bruhat-Tits tree for $\mathbb{Q}_2$.} \label{bruhat-tits-2}
\end{figure}

The $p$-adic field $\mathbb{Q}_p$ can be defined as the field of
Laurent series 
$$
\sum\limits_{\nu=-m}^\infty a_\nu p^\nu,\quad a_\nu\in\mathset{0,\dots,p-1}.
$$ 
It is well known that the $p$-adic norm induces a topology on the
 field $\mathbb{Q}_p$ which makes it into a totally disconnected space.
This is, however, compensated by the fact that $p$-adic discs never
overlap. Hence, the ultrametric inequality provides us with a tree-like
topology on the set of discs. It is precisely this hierarchical structure
of discs which makes $p$-adic numbers interesting for hierarchical classification. Consider the unit disc 
$$
\mathbb{D}=\mathset{x\in\mathbb{Q}_p\colon\absolute{x}_p\le 1}=B_1(0). 
$$
It is a subring of $\mathbb{Q}_p$ which coincides with the
ring of $p$-adic integers 
$$
\mathbb{Z}_p=\mathset{\sum\limits_{\nu=0}^\infty a_\nu p^\nu\colon a_\nu\in\mathset{0,\dots,p-1}}.
$$
It has a unique maximal ideal $p\mathbb{Z}_p$, and this ideal 
coincides with the maximal ``open'' (non-trivial) subdisc 
$\mathset{x\in\mathbb{Q}_p\mid\absolute{x}_p<1}$. It is a standard fact from
algebra that the quotient of a unital commutative ring by a maximal ideal is
a field. In our case, $\mathbb{Z}_p/p\mathbb{Z}_p\cong\mathbb{F}_p$,
the finite field with $p$ elements. This is well known and follows from the
fact that the unit disc
 is covered by the finite number of translates of the subdisc
$p\mathbb{Z}_p$:
$$
\mathbb{Z}_p=\bigcup_{x=0}^{p-1}(x+p\mathbb{Z}_p),
$$
which says in a fancy way that there are precisely
$p$ choices for the constant term in the power series expansion of
any $p$-adic integer. Hence, we have a hierarchical structure
of a disc with $p$ maximally smaller subdiscs. By rescaling and translation, it follows
immediately that any $p$-adic disc has precisely $p$ smaller subdiscs which are
maximal as subdiscs. Observe that $p\mathbb{Z}$ has precisely one minimal
bigger disc containing $p\mathbb{Z}_p$, namely $\mathbb{D}$. Again, this
holds for all $p$-adic discs. The consequence is that the set of all subdiscs
of $\mathbb{Q}_p$ form a $p+1$-regular tree $\mathscr{T}_{\mathbb{Q}_p}$,
 called the {\em Bruhat-Tits tree} for $\mathbb{Q}_p$.
Figure \ref{bruhat-tits-2} shows an illustration of
 $\mathscr{T}_{\mathbb{Q}_2}$ taken from \cite[Fig.\ 5]{CK2005}.

\subsubsection{Bruhat-Tits trees for $p$-adic number fields}

The field $\mathbb{R}$ of real numbers
is complete with respect to the archimedean distance
$\absolute{\cdot}_{\mathbb{R}}$. In the same way is
the field $\mathbb{Q}_p$  complete with respect to $\absolute{\cdot}_p$.
However, neither $\mathbb{R}$ nor $\mathbb{Q}_p$ is algebraically closed.
In the archimedean case, the algebraic closure of $\mathbb{R}$ is the field
$\mathbb{C}$ of complex numbers, and $\mathbb{C}$ is a two-dimensional
vector space over the scalar field $\mathbb{R}$.
By definition, the {\em degree} of a field extension $L$ over $K$ (meaning
$K$ is a subfield of a field $L$) is the dimension of $L$ as a vector space
over the scalar field $K$. If that degree is finite, then $L$ is called
a {\em finite extension} of $K$.
Hence, the degree of $\mathbb{C}$ over $\mathbb{R}$ is $2$, and $\mathbb{R}$ has no other finite field extensions.
In contrast, $\mathbb{Q}_p$ has extension fields of arbitrary degree.
Hence, the algebraic closure of $\mathbb{Q}_p$ is an infinite extension of $\mathbb{Q}_p$. Assume that
a finite extension field $K$ of $\mathbb{Q}_p$ of degree $n$ be given. Then it is known
that the distance $\absolute{\cdot}_p$ extends uniquely to a norm
$\absolute{\cdot}_K$ on $K$, and $K$ is complete with respect to $\absolute{\cdot}_K$ \cite[\S 5.3]{Gouvea}. Again the unit disc 
$$
\mathcal{O}_K=\mathset{x\in K\colon\absolute{x}_K\le 1}
$$
is a ring with unique maximal ideal 
$$
\mathfrak{m}_K=\mathset{x\in K\colon\absolute{x}_K<1},
$$
and $\kappa=\mathcal{O}_K/\mathfrak{m}_K$ is a finite field extension of $\mathbb{F}_p$, called the {\em residue field}.
It is finite with $p^f$ elements, if $f$ is the degree of $\kappa$
over $\mathbb{F}_p$. 
In general, the degree $n$ is not smaller than $f$, but if $n=f$, then
$K$ is called {\em unramified} over the subfield $\mathbb{Q}_p$.
A finite extension field of $\mathbb{Q}_p$ is also called a {\em $p$-adic number field}, and the elements of $\mathbb{Q}_p$ are sometimes called {\em rational
$p$-adic numbers}.

In any case, if $K$ is a $p$-adic number field 
with $\kappa\cong\mathbb{F}_{p^f}$, then in the same manner as 
with $\mathbb{Q}_p$ the unit disc $\mathcal{O}_K$ is covered by $p^f$ translates
of the subdisc $\mathfrak{m}_K$. This gives rise to the {\em Bruhat-Tits tree} 
$\mathscr{T}_K$ for $K$ which is an infinite $p^f+1$-regular tree. In other words, the number
of edges emanating from a vertex of $\mathscr{T}_K$ depends on the residue
field $\kappa$ which can be the same for different extension fields of 
$\mathbb{Q}_p$. So,  the choice of unramified extensions
is in some sense optimal for constructing the Bruhat-Tits trees. 

In general, $p$ will not be a prime element of $\mathcal{O}_K$, but
this is true in the unramified case. In fact, for $K$ unramified
of degree $f$ over 
$\mathbb{Q}_p$, it holds true that $\mathfrak{m}_K=p\mathcal{O}_K$,
and every element of $\mathcal{O}_K$ has an expansion
$$
x=\sum\limits_{\nu=0}^\infty a_\nu p^\nu,\quad a_\nu\in\mathfrak{R},
$$
where $\mathfrak{R}$ is a system of $p^f$ representatives modulo 
$p\mathcal{O}_K$.
However, in the {\em ramified} case, $p$ is not a prime in
 $\mathcal{O}_K$. But also in this  case, the maximal ideal 
$\mathfrak{m}_K$ is of the form $\pi\mathcal{O}_K$ for some prime $\pi\in\mathcal{O}_K$. It is always possible to choose $\pi$ such that $\absolute{\pi}_K=p^{-1/e}$ for some natural number $e\ge 1$ \cite[\S 5.4]{Gouvea}. This number $e$ is called the {\em ramification index}
of $K$ over $\mathbb{Q}_p$, and $K$ is {\em ramifed} over $\mathbb{Q}_p$,
if $e>1$. 
The  extension $K$ over $\mathbb{Q}_p$ is called
{\em purely ramified}, if  $f=1$.

\subsubsection{Cyclotomic $p$-adic fields} 

It is known that $\mathbb{Q}_p$ contains the $p-1$-st of unity,
but not the $p^f-1$-st  roots  of $1$ for $f>1$. 
Therefore, we discuss
the fields $\mathbb{Q}_p(\zeta)$ obtained by adjoining 
 to $\mathbb{Q}_p$ the $n$-th roots
of unity which are all powers of $\zeta$, a primitive $n$-th root 
of $1$. We will
first consider the case that $p$ is prime to $n$.
In that case, $\mathbb{Q}_p(\zeta)$ is unramified over $\mathbb{Q}_p$,
 the  degree is the smallest number $f$ such that $p^f\equiv 1\mod n$,
and
$\mathset{1,\zeta,\dots,\zeta^{f-1}}$ represents an $\mathbb{F}_p$-basis of 
$\kappa=\mathbb{F}_{p^f}$ in $\mathcal{O}_{\mathbb{Q}(\zeta)}$
which equals the polynomial ring $\mathbb{Z}_p[\zeta]$ 
\cite[II.(7.12)]{Neukirch}. That means, we can choose  
$$
\mathfrak{R}=\mathset{\sum\limits_{\nu=0}^{f-1}a_\nu\zeta^\nu\colon a_\nu\in\mathset{0,\dots,p-1}}
$$
as a system of representatives which is in bijection with a subset of
$\mathbb{N}^f$.

The most economic choice for $n$ is certainly
$p^f-1$. In that case, $\mathbb{Q}(\zeta)$ is again unramified of degree $f$
over $\mathbb{Q}_p$, and
$$
\mathfrak{R}_T=\mathset{0,1,\zeta,\dots,\zeta^{p^f-2}}
$$
is an alternative set of representatives 
 for $\kappa=\mathbb{F}_{p^f}$. This is a consequence of {\em Hensel's Lemma} \cite[Thm.\ 5.4.8]{Gouvea} (cf.\ \cite[\S 5.4]{Gouvea}. The elements of $\mathfrak{R}_T$ are called
the {\em Teichm\"uller representatives} of $\mathbb{F}_{p^f}$ and have the
characterising property that the residue class of $\zeta$ generates the
multiplicative group 
$\mathbb{F}_{p^f}^\times=\mathbb{F}_{p^f}\setminus\mathset{0}$.
For example, if $f=1$, then already $\mathbb{Q}_p$ contains the 
$p-1$-st roots of $1$ which form the Teichm\"uller representatives
in that case \cite[Cor.\ 4.3.8]{Gouvea}.

Another important case is when $\zeta$ is a primitive $p^m$-th root of unity.
Then $\mathbb{Q}_p(\zeta)$ is purely ramified over $\mathbb{Q}_p$
of degree $e=(p-1)p^{m-1}$ \cite[II.(7.13)]{Neukirch}.
  
\subsection{Algebraic $p$-adic dendrograms} \label{p-ad-dendro}

The reason for introducing the Bruhat-Tits tree $\mathscr{T}_K$ also for finite
extensions $K$ of $\mathbb{Q}_p$ is that the number of children of a vertex
can in principle be unbounded. This means that $K$ must be taken sufficiently
large in order for a dendrogram to be embeddabe into $\mathscr{T}_K$. 
In this subsection, we will effect the embedding, define $p$-adic dendrograms
and discuss these from a geometric perspective.

\subsubsection{Cyclotomic encoding} \label{cyclotomiccode}

Let $X=(T,v,D,\mu)$ be a projective dendrogram. By the {\em children}
$\Child(v)$
of a vertex $w\in T^0$ we mean the outgoing edges of $v$ in $T^1$ which are not
labelled $\infty$. Let 
$$
m=\max\mathset{\#\Child(w)\mid w\in T^0},
$$ 
and $f$ minimal such that $m\le p^f$. $K=\mathbb{Q}_p(\zeta)$ 
will denote the cyclotomic field with $\zeta$ a primitive $p^f$-th 
root of unity, and assume that $\mathfrak{R}$ is a full
system of representatives modulo $p\mathcal{O}_K$ containing $0$ and $1$.
Generalising the binary case, $\chi_w\colon\Child(w)\to\mathfrak{R}$
is now an inclusion map for every vertex $w\in T^0$ such that $0\in\Image\chi_w$,
and $1\in\Image\chi_v$. These maps form together a map $\chi\colon T^1\to\mathfrak{R}$ which yields a $p$-adic encoding map
$$
\mathcal{G}(\infty,D)\to K,\; \gamma_x\mapsto\sum\limits_{\gamma_x}\chi(e)p^{\ell(o(e))},
$$
where $\ell\colon T^0\to\mathbb{N}$ is the level map derived from the metric $\mu$ as in Section \ref{abstractdendro}. Again, as in the binary case, the natural
identification $D\cong\mathcal{G}(\infty,D)$ yields a $p$-adic encoding
$\chi\colon D\to K$
of the data. In the case $m=2$, we recover the binary encoding as in Section \ref{binary} for ordererd dendrograms, if the local encoding maps $\chi_w\colon\Child(w)\to\mathfrak{R}=\mathset{0,1}$ are
chosen appropriately.

\begin{Rem} 
If a dendrogram is binary, or the prime $p$ is sufficiently large (not smaller than the largest number of children of any given vertex), then a rational $p$-adic encoding is possible. In this case, data will be represented by finite $p$-adic expansions, hence by natural numbers.  
Restricting to {\em rational} $p$-adic encoding has the disadvantage that  
$p$ is a fixed bound for the number of possible children vertices in dendrograms. Hence, if there is no a priori bound in data, then it is necessary to  
allow unramified extensions  of $\mathbb{Q}_p$ of arbitrary degree.  
From a computational point of view it is probably most interesting 
to keep the prime $p$ as low as possible, i.e.\ $p=2$.   
\end{Rem} 

\subsubsection{Dendrograms and the $p$-adic projective line} \label{*-tree}

Let $X$ be a projective dendrogram. In Section \ref{cyclotomiccode},
we have constructed an embedding of the underlying data $D\to K$
into a $p$-adic number field $K$. From a geometric viewpoint,
this is an embedding of $D$ into the $p$-adic affine line.
It is often 
 convenient to treat points on the affine line and
$\infty$  on an equal footing. The
geometric space  enabling  this is the {\em projective line}.
Hence, we consider a $p$-adic number from $K$ as a point
in the $p$-adic projective line $\mathbb{P}^1$. 
The space $\mathbb{P}^1$ is a $p$-adic manifold defined over $\mathbb{Q}_p$
and
 whose $K$-rational
points are given by $\mathbb{P}^1(K)=K\cup\mathset{\infty}$ for any field $K$
containing $\mathbb{Q}_p$. Here, we consider only the case that $K$ 
is a $p$-adic number field. 

In Section \ref{btt}, we have constructed for each $K$  the Bruhat-Tits tree
$\mathscr{T}_K$ by associating to each disc in $K$ a vertex of $\mathscr{T}_K$.
To an inclusion $B\subseteq B'$ of disks corresponds a geodesic path
 between the associated vertices $w$ and $w'$. It is a fact that any 
strictly descending infinite chain of discs in $K$
\begin{align}
B_1\supseteq B_2\supseteq\dots \label{chain}
\end{align}
converges to a $K$-rational point: $\mathset{x}=\bigcap B_\nu$
with $x\in\mathbb{P}^1(K)$. In the tree $\mathscr{T}_K$,
the chain (\ref{chain}) corresponds to an infinite geodesic half-line
$$
\xymatrix{
*\txt{$\stackrel{v_1}{\bullet}$}\ar@{-}@<-3pt>[r]
&*\txt{$\stackrel{v_2}{\bullet}$}\ar@<-3pt>@{-}[r]&\stackrel{\rule{0pt}{5pt}}{\dots}
}
$$
and the $p$-adic number $x$ lies at its end. It is a well known fact
that the ends of $\mathscr{T}_K$ correspond bijectively to $\mathbb{P}^1(K)$.
Now let $S\subseteq\mathbb{P}^1(K)$ be a finite set of $p$-adic numbers.
Then we can form the {\em $*$-tree} which is the
 smallest subtree $T^*\langle S\rangle$ of $\mathscr{T}_K$
whose ends correspond to $S$. 
If $S\setminus\mathset{\infty}\subseteq\mathscr{O}_K$  and
$S$ contains $0$ and $1$, then $T^*\langle S\rangle$ 
can be made in a natural way to a projective dendrogram. 
First observe that three disctinct points $x,y,z\in\mathbb{P}^1(K)$
define a unique vertex $v(x,y,z)$ in $\mathscr{T}_K$: it is the
intersection of the three geodesics ending in $\mathset{x,y,z}$.
Hence, $T^*\langle S\rangle$ contains the vertex $v=v(0,1,\infty)$
corresponding to the unit disc $\mathcal{O}_K$. Choose $v$ as the root. 
As  $S\subseteq\mathcal{O}_K\cup\mathset{\infty}$, all vertices 
on the
half-line $]v,\infty[$ have precisely two emanating edges in 
$T^*\langle S\rangle$.
By defining $T^0_S$ to be the set 
of vertices $w$ in $T^*\langle S\rangle$ with $\#\Child(w)\ge 2$,
and $T^1_S$ the set of geodesic paths $]w,w'[$
between  $w\in T^0_S$ and $w'\in T^0_S\cup S$
not containing a vertex from $T^0_S$, we obtain a tree $T_S$
whose data $D_S$ are the half-lines $]w,s[$ with $w\in T^0_S$, $w\in S\setminus\mathset{\infty}$ and not containing any vertex from $T^0_S$. 
This yields a projective
dendrogram $X_S=(T_S,v,D_S,\mu_S)$,
where the metric $\mu$ is defined as
 the number 
$\mu(e)$ of edges in $\mathscr{T}_K$ on the geodesic path corresponding to
$e\in T^1_S$.
It is clear that there is a natural $p$-adic encoding $\chi_S\colon D_S\to S$
with numbers from $K$. 

A tree in which every vertex $v$ has more than two emanating edges
is called {\em stable}. This is, in a way, a kind of minimal
representation of a tree.
In that sense, $T_S$ is the {\em stabilisation}
of $T^*\langle S\rangle$. 
It is clear that for any projective dendrogram
$X=(T,v,D,\mu)$ a $p$-adic endoding $\chi\colon D\to K$ as in Section \ref{cyclotomiccode} yields a tree $T^*\langle\chi(D)\rangle$ whose stabilisation
is tree-isomorphic to $T$. 
Hence, a $p$-adic encoding $\chi$ 
means in $p$-adic geometry
an embedding of projective
 dendrograms into $\mathscr{T}_K$
for some $p$-adic field $K$ through the assignment
$$
\Xi\colon X\mapsto \mathbb{P}^1\setminus(\chi(D)\cup\mathset{\infty}). 
$$
This assignment is a map from the space 
$\mathfrak{D}_n$ of  dendrograms on $n$ data
to the space $\mathfrak{M}_{0,n+1}$ of $n+1$-pointed projective lines.
Any pointed projective line is, by means of a projective
linear transformation,
represented by 
$\mathbb{P}^1\setminus\mathset{x_0,\dots,x_n}$ 
such that $x_0=0$, $x_1=1$, $x_2=\infty$.  

\begin{Def}
A {\em $p$-adic dendrogram} is a pair $(X,\chi)$  with  a projective
dendrogram $X$ and a map $\chi\colon D\to K$
into
 some $p$-adic number field $K$ such that there is an isometric isomorphism
between ${T}^*\langle\Image(\chi)\cup\mathset{\infty}\rangle$ and 
the underlying tree of $X$. A $p$-adic dendrogram is called {\em normal}
if $0,1\in\Image(\chi)\subseteq\mathcal{O}_K$. 
\end{Def}

\subsubsection{Binary data are generic}

The space $\mathfrak{D}_{n}$ is known to be a polyhedral complex of
dimension $n-2$, and the cells  of maximal dimension consist of the
dendrograms whose underlying trees are binary. 
In fact, the dimension equals the number of internal edges (which can be
of arbitrary length), and for binary dendrograms this number is $n-2$.  
The other dendrograms
are all contained in lower dimensional cells.
As the latter are obtained by contracting edges of binary dendrograms,
the corresponding cells in $\mathfrak{D}_n$ are always in the boundary
of cells of maximal dimension $n-2$. Hence, binary dendrograms are generic.

The $*$-tree construction from Section \ref{*-tree} gives a map
$$
\Theta\colon\mathfrak{M}_{0,n+1}\to\mathfrak{D}_n,\;
\mathbb{P}^1\setminus\mathscr{L}\mapsto T^*\langle\mathscr{L}\rangle
$$
where $\mathscr{L}\subset\mathbb{P}^1(K)$ is assumed to contain $0,1,\infty$.
This map  is the {\em Tate map} and is many-to-one. In fact, its 
fibres  are open in the analytic
topology of $\mathfrak{M}_{0,n+1}$. The relation to 
the above is that 
 the $p$-adic encoding map $\Xi\colon\mathfrak{D}_n\to\mathfrak{M}_{0,n+1}$
is a section of the Tate map, i.e.\
$\Theta\circ\Xi=\Identity_{\mathfrak{D}_n}$.

From this geometric point of view, a dendrogram is merely a point in
$\mathfrak{D}_n$, and a $p$-adic dendrogram is determined by
a point  $x\in\mathfrak{M}_{0,n+1}$. A time series of dendrograms 
is given as a map $\mathset{x_0,\dots,x_N}\to\mathfrak{D}_n$
or, in the $p$-adic case, $\mathset{x_0,\dots,x_N}\to\mathfrak{M}_{0,n+1}$.

In what follows, we consider w.l.o.g.\ $p$-adic dendrograms.
In general, a {\em family of dendrograms with $n$ data} is given by a map 
$S\to\mathfrak{M}_{0,n+1}$ for some $p$-adic space $S$.
By $p$-adic geometry, there is an associated
continuous map $\Sigma\to\mathfrak{D}_n$
for some real space $\Sigma$ depending on $S$. The space $S$
is viewed as a {\em parameter space} for the family: small variations
in $S$ yield nearby dendrograms.
Hence, we can  speak of a {\em small deformation} of dendrograms:
this is a family of dendrograms such that $\Sigma\to\mathfrak{D}_n$
maps into a fixed cell. In that case, the topological type
of each dendrogram parametrised by $S$ (or $\Sigma$) is always the same,
only the lengths of internal edges vary in the family.
More details on families of dendrograms can be found in \cite{BradDegendendrofam,BradDendrofam}.

\section{Classification of strings}

By using a finite unramified extension $K$ of $\mathbb{Q}_p$,
one obtains an encoding of data by finite expansions in powers of $p$
and coefficients in a system $\mathfrak{R}$ of representatives
modulo $p\mathcal{O}_K$. We will always assume that $\mathfrak{R}$
contains $0$ and $1$, so that we can then speak of
polynomials in $p$ over $\mathfrak{R}$. We denote by $\mathfrak{R}[p]$ the
 set of all such polynomials. 

\subsection{Cyclotomic $p$-adic encoding of strings} 
 
Let $\mathcal{A}$ be some finite alphabet, and $S(\mathcal{A})$ the set of all possible strings using letters from $\mathcal{A}$. In other words,  
$S(\mathcal{A})$ is the set of infinite 
sequences of letters from $\mathcal{A}$.
We will interpret finite strings also as infinite sequences by assuming
that $\mathcal{A}$ contain a distinguished letter $0$ or ``blank'',
and a string is {\em finite}, if only finitely many of its letters are 
not blank. The set of finite strings is denoted by $S_{\Finite}(\mathcal{A})$.
$S(\mathcal{A})$ is endowed with the ultrametric
{\em Baire distance}
$$
\delta_p\colon S(\mathcal{A})\times S(\mathcal{A}) \to\mathbb{R},\quad
(x,y)\mapsto\inf\mathset{p^{-n}\mid x[n]=y[n]}
$$
where $z[n]$ denotes the sequence of the first $n$ letters in the string $z$.
Usually, $\delta_2$ is used as the Baire distance.
It is an ultrametric which resembles very much the $p$-adic distance.
In any case, it is an easy exercise to prove that $(S(\mathcal{A}),\delta_p)$
is a complete metric space and that $S_\Finite(\mathcal{A})$ is a dense
subspace
of $S(\mathcal{A})$.
 
\begin{thm} \label{pi-adic-strings} 
There exists a $p$-adic number field  $K$ unramified of degree 
$f$ over $\mathbb{Q}_p$,  
 a full system $\mathfrak{R}\subseteq\mathcal{O}_K$  
 of representatives modulo $p\mathcal{O}_K$, and a closed isometric 
embedding
$\phi\colon (S(\mathcal{A}),\delta_p)\to(\mathcal{O}_K,\absolute{\cdot}_K)$ 
which
takes $S_\Finite(\mathcal{A})$ into 
$\mathfrak{R}[p]\subseteq\mathcal{O}_K$.  The set
 $\phi(S_\Finite(\mathcal{A}))$  is dense in $\Image(\phi)$.
\end{thm} 
 
\begin{proof} 
Take $f$ sufficiently large, and idenfify $\mathcal{A}$ with a subset of $\mathfrak{R}$ in such a way that the blank maps to $0\in\mathfrak{R}$. Clearly,
the distances coincide after identification, and the statements of the
theorem follow from this. 
\end{proof} 
 
\begin{Rem}
The isometric map $\phi$ in Theorem \ref{pi-adic-strings} identifies
$S(\mathcal{A})$ with a so-called {\em affinoid disc}, i.e.\
a closed disc with ``holes''. In fact, $\Image(\phi)$ is the
unit disc $\mathcal{O}_K$ minus the preimage of some points of $\kappa$ under
the canonical projection 
$\rho\colon\mathcal{O}_K\to\mathcal{O}_p/p\mathcal{O}_p=\kappa$.
\end{Rem}

\begin{Exa} 
Consider the strings in the letters $\mathset{A,G,C,T}$ representing DNA sequences in the four nucleotides adenine $(A)$, guanine $(G)$, cytosine $(C)$ and thymine $(T)$. In \cite{DNAp} a rational $5$-adic model for such strings is discussed, and combined with a $2$-adic distance. 
The encoding in \cite{DNAp} identifies the nucleotides with 
$\mathset{1,2,3,4}$.
Hence, the code alphabet is $\mathcal{A}=\mathset{0,1,2,3,4}$,
and the finite rational $5$-adic numbers represent all
finite lists of 
nucleotides with arbitrarily long spacings between them.
 
We show that one could use a model based on a single prime, namely $p=2$, using an extension field $K$ finite over $\mathbb{Q}_2$.  
 
As we are using four letters, the $2$-adic field $\mathbb{Q}_2$ is too small, 
because its residue field $\mathbb{F}_2$ has only 2 elements. However, 
a $2$-adic field $K$ with residue field $\kappa\cong\mathbb{F}_{2^2}$ would be precisely sufficient, if we do not care 
about blanks. 
This can be realised thus: take a primitive third
 root $\zeta$ of unity, and let $K=\mathbb{Q}_2(\zeta)$ be the corresponding cyclotomic field extension. By number theory, $K$ is unramified of degree $f=2$ over $\mathbb{Q}_2$, 
because $2^f\equiv 1\mod 3$, and $f=2$ is minimal with that property. 
 
As $K$ is unramified over $\mathbb{Q}_2$, $2$ is a prime of 
$\mathcal{O}_K\cong\mathbb{Z}_2[\zeta]$. 
Then $\mathfrak{R}_T=\mathset{0,1,\zeta,\zeta^2}$ is 
the system of Teichm\"uller
 representatives for $\mathcal{O}_K/2\mathcal{O}_K\cong\mathbb{F}_{2^2}$, and we have 
$$ 
\mathcal{O}_K=\mathset{\sum\limits_{\nu=0}^\infty a_\nu 2^\nu\colon a_\nu\in\mathfrak{R}_T}. 
$$  
Now, any bijection $\mathset{A,G,C,T}\cong\mathfrak{R}_T$ yields a $2$-adic encoding of DNA. 
However, this method does not distinguish between $0$ and blank,
so it is never clear, how long a string represented by a $2$-adic number 
is supposed to be.
On the other hand,
there is already an existing proposal in \cite{KK2007}
 based on the single prime $2$. There, the bijection
\begin{align*}
\mathset{A,G,T,C}&\to\mathbb{F}_2^2,\\
A\mapsto (0,0),\;&
G\mapsto (0,1),\; T\mapsto (1,0),\;C\mapsto(1,1)
\end{align*}
is proposed.
If we take the isomorphism $\mathbb{F}_2^2\cong\mathbb{F}_{2^2}$
defined by $(1,0)\mapsto 1$, $(0,1)\mapsto\bar\zeta$,
where $\bar\zeta\in\mathbb{F}_{2^2}$ is the residue class 
of $\zeta$,
this amounts to encoding DNA by $\mathfrak{R}=\mathset{0,1,\zeta,1+\zeta}$,
and we obtain the bijections of ordered sets
$$
(A,G,T,C)\cong(0,\zeta,1,1+\zeta)\cong(0,1,\zeta,\zeta^2).
$$
The authors of \cite{KK2007} consider only words of fixed length $3$,
wherefore the question ``$0=\text{blank}$?'' does not arise there. 

In any case, if we take $f=3$, then $\mathbb{Q}_2(\zeta)$ is certainly
large enough to include ``blank'' in our $2$-adic alphabet for DNA.  
\end{Exa} 

Next, we observe that cyclotomic encoding is persistent:
 
\begin{thm} \label{cyclotomicstrings}
Every finite alphabet has a cyclotomic encoding for every prime $p$. 
\end{thm} 
 
\begin{proof} 
We need to show that for all $f$ there is a natural number $n$ 
such that $f$ is minimal with $p^f\equiv 1\mod n$. Taking $n=p^f-1$
sufficiently large proves the assertion. 
\end{proof} 

\begin{Rem}
Note that arbitrary sets of strings form dendrograms for the Baire distance
which in general are not normal. In the finite 
case, it is possible to
make the dendrogram normal by a shift (which corresponds to multiplication
by some $p$-adic integer).
\end{Rem}

\begin{Rem}
The authors of \cite{MDC2006} consider a variant of the Baire distance
on strings.
Namely, for $k\ge 1$ let 
$$
d_k(x,y)=\inf\mathset{p^{-n}\mid x[n]=y[n],\; 0\le n\le k}
$$
which they consider for $p=2$. This distance does not distinguish
between strings with a common prefix of length $k$ or more. Equivalently,
the corresponding $p$-adic numbers are not distinguished, if
their $p$-adic distance equals $d_k(x,y)$ or less. 
\end{Rem}

\subsection{A hierarchic algorithm for strings}

The main advantage of strings is that, by
Theorem \ref{cyclotomicstrings}, the extension field
$K$ can be a priori chosen as a cyclotomic field of fixed degree, 
as it is determined by the size of the alphabet.

\subsubsection{General description}

A solely $p$-adic agglomerative hierarchic algorithm for strings
is now  presented which does without the changing of 
the distance function usual in the archimedean case. The reason is
that by the ultrametric triangle inequality, the distance between two
disjoint discs $B$ and $B'$
 equals the distance between any two representatives $x\in B$ and $x'\in B'$.
It can be essentially broken  down into two steps.

\medskip
Step 1. Encode strings by $p$-adic numbers from the cyclotomic field $K$.

\medskip
Step 2. Classify $p$-adic numbers using  $\absolute{\cdot}_K$.

\medskip
Step 1 has been described in the previous subsection, and Step 2 will
be explained below.
The output is the uniquely determined $*$-tree for the given strings.

\begin{Rem}
The algorithm in Step 2 is independent of whether the $p$-adic numbers
 encode strings or not. In fact, it merely classifies $p$-adic numbers.
Hence, the focus in $p$-adic hierarchical classification of data
which are not to be taken as strings lies in
the analogue of Step 1 which is unsolved as far
as the author is aware.
\end{Rem}

\subsubsection{The classification algorithm}
Let $D=\mathset{x_1,\dots,x_n}$ be a set of $n$ different $p$-adic numbers.
We assume that these are taken from some cyclotomic $p$-adic field 
$K=\mathbb{Q}_p(\zeta)$. In the special case $K=\mathbb{Q}_p$,
encoding by natural numbers might be tempting. Then the euclidean
algorithm will yield the $p$-adic expansion. So, we assume that
all $x\in D$ are given by their $p$-adic expansion
with coefficients in a full system $\mathfrak{R}$ of
representatives modulo $pO_K$.

First note, that the computation of $\absolute{x}_K$ is simple,
if $x$ is given by the $p$-adic expansion. 

1. Take all $x\in D\setminus \mathset{x_1}$ with $\absolute{x_1-x}_K$ 
   minimal to form the cluster $C(x_1)$. Do the same with all other $x_i\in D$
and obtain the clusters $C(x_1),\dots,C(x_n)$ together with all possible
inclusions among these clusters.

2. Let $D'$ be the set of all maximal clusters among the $C(x)$ from 1.
Proceed with $D'$ in the same way as with $D$ in 1.\ using the 
$p$-adic distance between clusters. It is, by ultrametricity, 
given by $\absolute{x-y}_K$
for any points $x,y$ representing the clusters. Obtain $m<n$ clusters
and their hierarchy.

3. Let $D''$ be the set of all maximal clusters among the ones obtained in 2.
Etc.

As in each step, the number of clusters obtained strictly decreases, the
algorithm terminates with one single cluster. This must be $D$,
as otherwise one would go on  clustering. Putting togeter all hierarchies
yields the tree $T^*(D)$, at least topologically.
Taking some extra care in each step, yields the metric or level structure
on $T^*(D)$, as can easily be seen. 

\begin{Rem}
Clearly, for a given set $\mathcal{S}$ of strings, the output $T^*(D)$ depends on 
the encoding 
$\chi\colon\mathcal{S}\to D\subseteq\mathfrak{R}[p]$
of the strings. So, if $\mathfrak{R}$ is replaced by a
different set $\mathfrak{R}'$ modulo $p\mathcal{O}_K$,
then we obtain another encoding $\chi'\colon\mathcal{S}\to\mathfrak{R}'[p]$
by composing with any bijection $\phi\colon\mathfrak{R}\to\mathfrak{R}'$:
$\chi'=\phi\circ\chi$. Assume that the change of coefficients $\phi$
is such that it takes any element $a\in\mathfrak{R}$ 
representing $\bar{a}\in\kappa$ to $\phi(a)\in\mathfrak{R}'$ representing the same
$\bar{a}$ in the residue field $\kappa$, i.e.\
there is a commutative diagram
$$
\xymatrix{
\mathfrak{R}\ar[rr]^\phi\ar[dr]_{\rho_{\mathfrak{R}}}&
&\mathfrak{R}'\ar[dl]^{\rho_{\mathfrak{R}'}}\\
&\kappa&
}
$$
where $\rho_{\mathfrak{R}}$ and $\rho_{\mathfrak{R}'}$
are the restrictions of the canonical projection $\rho\colon\mathcal{O}_K\to\kappa$ to $\mathfrak{R}$ and $\mathfrak{R}'$, respectively.
Then the corresponding
$*$-trees are isometric, hence yield the same dendrograms.
\end{Rem}

\section{Discrete symmetries of time-series} 
 
In a time series of dendrograms $X_t$
with fixed number of ends, we consider the underlying data $D_t$
as a set of {\em particles} which ``move'' with respect to another
in ``time'' $t$, e.g.\ by using the same set of labels
for all data $D_t$. Naturally, a series of lists of strings can be 
considered as such a time series. 
Fixing the data size means that we assume there to be no collisions
among particles (cf.\ \cite{BradDegendendrofam} for colliding particles). 
Another technical simplification we make is that we consider
only binary dendrograms. The general case will be treated elsewhere.

\subsection{The $\dagger$-tree} 

\begin{figure}
$$
\xymatrix@=8pt{
&&*\txt{$\stackrel{v}{\bullet}$}\ar@{-}[dr]^1\ar@{-}[ddl]_2&\\
&&&*\txt{$\bullet$}\\
&*\txt{$\bullet$}\ar@{-}[dr]^1\ar@{-}[dddl]_3&&\\
&&*\txt{$\bullet$}\ar@{-}[dr]^1&\\
&&&*\txt{$\bullet$}\\
*\txt{$\bullet$}\ar@{-}[dr]^1&&&\\
&*\txt{$\bullet$}&&
}
$$
\caption{The $\dagger$-tree of the dendrogram in Figure \ref{Murtaghdendro}.}
\label{daggertree}
\end{figure}
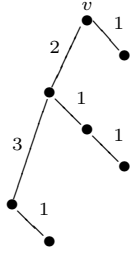

The tree $T$ underlying a dendrogram $X$ has an important subtree which depends
on the data: namely, the subtree $T^\dagger$ spanned by the vertices of $T$.
Its $p$-adic counterpart is the subtree $T^\dagger\langle S\rangle$
 of $T^*\langle S\rangle$ spanned by the vertices $v(x,y,z)$ where
$x,y,z\in S\subseteq\mathbb{P}^1(K)$ are three distinct points. 
We will also speak of a $\dagger$-tree when meaning $T^\dagger$.
For example,
the $\dagger$-tree of the dendrogram in Figure \ref{projdendro} is a segment
$\xymatrix{*\txt{$\bullet$}\ar@{-}[r]&*\txt{$\bullet$}}$, and $T^\dagger$
for  the tree $T$ underlying the dendrogram in Figure \ref{Murtaghdendro}
is shown in Figure \ref{daggertree}, where the numbers indicate 
the edge lengths.

The $\dagger$-tree gives a rough idea on the distribution of the 
distances within data. We define for this end the {\em volume}
of $T^\dagger$ (or a dendrogram $X=(T,v,\lambda,D,\mu)$) as the total
length of its edges:
$$
\Vol(T^\dagger)=\sum\limits_{e\in T^{\dagger,1}}\mu(e).
$$
Each child $e$ of $v$ gives rise to a branch $\Gamma$ consisting
of  all geodesic paths beginning in the target vertex of $e$
and directed away from $v$. The {\em weight} of branch $\Gamma$
is now defined as
$$
w(\Gamma)=\Vol(\Gamma)+\mu(e)=\Vol(\Gamma^\dagger)+\mu(e).
$$
The influence on the dendrogram $X$ is measured by 
a complex number we call the
{\em balance} of $X$:
$$
b(X)=\sum\limits_{\nu=0}^{m-1}w_\nu e^{\frac{2\pi\sqrt{-1}}{m}}\in\mathbb{C},
$$
where $\Gamma_0,\dots,\Gamma_{m-1}$ are the different branches
and $w_\nu=w(\Gamma_\nu)$. 
$X$ is {\em balanced}, if $b(X)=0$. This occurs if and only if all weights
$w_\nu$ are equal. 

\begin{Exa}
The dendrogram $X$ in Figure \ref{Murtaghdendro} has the following
values for the quantities:
$$
\Vol(X)=9,\; w(\Gamma_0)=8,\; w(\Gamma_1)=1,\; b(X)=7.
$$
\end{Exa}

Studying the balance $b(X_t)$ of a time series $X_t$
 gives a first indication
on the behaviour of the growth of individual branches.

\begin{Rem}
The $\dagger$-tree of a $p$-adic dendrogram indicates the amount
of freedom one has for the coding map $\chi$. Namely, data $D$ can be 
given any $p$-adic values $\mathscr{L}$ or $\mathscr{L}'$, 
as long as 
$$
T^\dagger\langle\mathscr{L}\cup\mathset{\infty}\rangle 
=T^\dagger\langle\mathscr{L}'\cup\mathset{\infty}\rangle
$$
holds true.
This means for the $p$-adic expansions 
that the coefficients of the high powers of $p$
can be chosen arbitrarily from $\mathfrak{R}$.
\end{Rem}

\subsection{Time-invariant subtrees of $p$-adic dendrograms}

Any edge $e$ in the underlying rooted tree $(T,v)$ of a dendrogram $X$
defines a {\em branch} $\Gamma_e$. It is itself a dendrogram 
with data $D_e$ and
is the union of $e$ and
the subtree of $T$ spanned by all vertices of $T$ below $e$. 
In order to be able to compare the evolution in time of dendrograms,
we assume that we are given a family 
$F\colon\mathset{X(0),\dots,X(N)}\to\mathfrak{M}_{0,n+1}$ of normal $p$-adic
dendrograms with $n$ data. In this case, $v$ is a fixed vertex of the time
series $F$. This time seris $F$ defines a family of subtrees 
$T^*(i)=T^*\langle \chi(D_i)\cup\mathset{\infty}\rangle$ of the Bruhat-Tits tree $\mathscr{T}_K$
for $K$ a sufficiently large $p$-adic number field,
where $\chi$ is the coding map associated to $F$, and $D_i$ the data of $X(i)$.
A subtree $\Gamma$ of some $T^\dagger(i)$ is 
said to be {\em time-invariant}, if $\Gamma$ lies in all $T^\dagger(i)$.
A geodesic $\gamma=]a,b[$ is {\em time-invariant}, if $\gamma$
lies in all $T^*(i)$ and at all times $(a,b)$ represents the same pair
of particles.
A branch $\Gamma$ of some $T^*(i)$
 is {\em time-invariant},
if the path from $v$ to the root $v_\Gamma$ of $\Gamma$ is
a time-invariant subtree, and the data 
adherent to $v_\Gamma$ by paths away from $v$
represent  the same set of particles at all times $t=0,\dots,N$.

\subsection{Time series of genus one} \label{time4Tate}

The definition of balance uses as point of reference the vertex $v$.
In our considerations, it will play the role of a ``fixed star''.

Assume for convenience 
that we are given a binary $p$-adic time series, that is a
time series  of binary dendrograms $X_0,\dots,X_N$
encoded in some $p$-adic number field $K$ unramified over $\mathbb{Q}_p$.
In the binary case, the balance of each $X_t$ is given as
$$
b(X_t)=w_0-w_1\in\mathbb{Z},\quad t=0,\dots,N.
$$

The intersection $T_t^\dagger\cap]0,1[$ is a segment $I_t=[v_0(t),v_1(t)]$
and contains $v$. 
Assume that $b(I_t)$ follows a linear trend with rational slope
$c=\frac{d}{e}$. If $c\neq 0$, we
may assume that  $d\in\mathbb{Z}$, $e\in\mathbb{N}$ and have no
common divisor. 
We call $c$ the {\em velocity} of the time series
$X_t$ along the geodesic path $]1,0[$. 

Consider w.l.o.g.\ the case $c<0$. This means that there is a net
flow of balance towards $1$.

\smallskip\noindent
{\bf Case $\forall t\colon v\in I_t^\circ=]v_0(t),v_1(t)[$.} 
This case will be called {\em flow from infinity} and can be interpreted
as there being a balance flow from outside the data.
This case will not be considered, although technically it should
be similar to the following case.

\smallskip\noindent
{\bf Case $\exists t_0 \;\forall t\ge t_0\colon v=v_0(t)$.}
Then $X_t$ follows a translation
$\tau$ along $]0,1[$ with velocity $c$, and $0$ is the repellant
fixed point of $\tau$, and $1$ the attracting fixed point.
If $e>1$, then $\tau$ does not act on the tree $\mathscr{T}_K$,
because translations on $\mathscr{T}_K$ can be only by multiple 
shifts of edges from $\mathscr{T}_K$. However, if $L$ is a $p$-adic 
number field which is {\em purely ramified} over $K$ with {\em ramification index} $e$, then there is a Bruhat-Tits tree $\mathscr{T}_L$ which is of the same
regularity as $\mathscr{T}_K$, and which topologically contains 
$\mathscr{T}_K$,
 but in which every edge of $\mathscr{T}_K$
is subdivided into $e$ edges of equal length $\frac{1}{e}$.
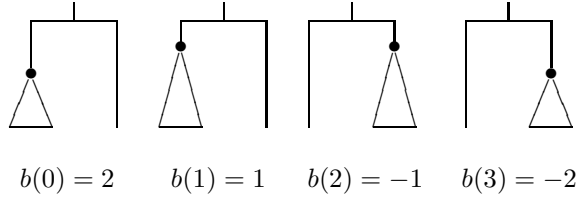
\begin{figure}
\centering
$
\begin{array}{cccc}
\xymatrix@!0@R=10pt@C=8pt{
&&&&\\
&*\txt{}\ar@{-}[dd]&&*\txt{}\ar@{-}[u]&*\txt{}\ar@{-}[r]\ar@{-}[lll]
  &*\txt{}\ar@{-}[dddd]\\
&&&&&\\
&*\txt{$\bullet$}\ar@{-}[ddl]\ar@{-}[ddr]&&&&\\
&&&&&\\
*\txt{}\ar@{-}[rr]&&*\txt{}&&&*\txt{}\\
&&&&&\\
&&&\hspace*{-5pt}b(0)=2&&
}
&
\xymatrix@!0@R=10pt@C=8pt{
&&&&\\
&*\txt{}\ar@{-}[d]&&*\txt{}\ar@{-}[u]&*\txt{}\ar@{-}[r]\ar@{-}[lll]
  &*\txt{}\ar@{-}[dddd]\\
&*\txt{$\bullet$}\ar@{-}[dddl]\ar@{-}[dddr]&&&&\\
&&&&&\\
&&&&&\\
*\txt{}\ar@{-}[rr]&&*\txt{}&&&*\txt{}\\
&&&&&\\
&&&\hspace*{-4pt}b(1)=1&&
}
&
\xymatrix@!0@R=10pt@C=8pt{
&&&&\\
*\txt{}\ar@{-}[dddd]\ar@{-}[rrrr]&&*\txt{}\ar@{-}[u]&&*\txt{}\ar@{-}[d]&\\
&&&&*\txt{$\bullet$}\ar@{-}[dddr]\ar@{-}[dddl]&\\
&&&&&\\
&&&&&\\
*\txt{}&&&*\txt{}\ar@{-}[rr]&&*\txt{} \\
&&&&&\\
&&&\hspace*{-6pt}b(2)=-1&&
}
&
\xymatrix@!0@R=10pt@C=8pt{
&&&&\\
*\txt{}\ar@{-}[dddd]\ar@{-}[rrrr]&&*\txt{}\ar@{-}[u]&&*\txt{}\ar@{-}[dd]&\\
&&&&&\\
&&&&*\txt{$\bullet$}\ar@{-}[ddr]\ar@{-}[ddl]&\\
&&&&&\\
*\txt{}&&&*\txt{}\ar@{-}[rr]&&*\txt{} \\
&&&&&\\
&&&\hspace*{-8pt}b(3)=-2&&
}
\end{array}
$
\caption{A time series of dendrograms.} \label{dendroseries}
\end{figure}

Note, that an extension of $p$-adic number fields
$L$ over $K$ is ramified, if  there is a prime $\pi_L$ of $\mathcal{O}_L$
such that for some $e>1$ holds true: $\absolute{\pi_L}_L^e=\absolute{\pi_K}$, 
where $\pi_K$ is a
prime of $\mathcal{O}_K$. The number $e$ is the ramification index.
The extension is {\em purely ramifed}, if the corresponding
extension of residue fields $\kappa_L$ over $\kappa_K$ has degree one.
 If $K$ is unramified over $\mathbb{Q}_p$, then 
it follows in any case that $\absolute{\pi_L}_L=p^{\frac{1}{e}}$.

Assume now that $c=\frac{d}{e}\neq 0$ with $e\ge 1$, and $d$ prime to $e$.
Let $L$ be a $p$-adic number field which is purely ramified over $K$ 
with ramification index $e$. Let $c_p\in L$ be such that
$\absolute{c_p}_L=p^{c}\neq 1$ 
(e.g.\ $c_p=p^{c}$, if this is an element of $L$).
Then the translation $\tau$ can be represented by the hyperbolic transformation
$\theta\colon z\mapsto\frac{-c_p}{(1-c_p)z-1}$ in the projective linear group
 $\PGL_2(L)$. This transformation $\theta$ can in turn be represented
by the matrix
$$
\Theta=\begin{pmatrix}-c_p&0\\1-c_p&-1\end{pmatrix}
$$
Because $\theta$ is hyperbolic, it generates a discrete subgroup 
$H=\langle\theta\rangle$ of $\PGL_2(L)$ which acts on the $p$-adic
manifold 
$\mathbb{G}_m=\mathbb{P}^1\setminus\mathset{0,1}$.
The quotient $E=\mathbb{G}_m/H$ is defined over $L$ and
known as a {\em Tate
curve}, i.e.\ a compact $p$-adic riemann surface of genus $1$,
i.e.\  the $p$-adic analogon of the surface of a torus.
Its $L$-rational points are given as
$$
E(L)=L^\times/H\cong \Ends(\mathscr{T}_L/H),
$$
where $\Ends(\mathcal{G})$ denotes the set of ends of a graph $\mathcal{G}$.
The quotient graph $\mathcal{E}=]0,1[\:/H$ is a loop, i.e.\ has 
first Betti number $\beta_1=h_1(\absolute{\mathcal{E}},\mathbb{R})=1$. And
the time series $X_t$ induces a dynamical system of vertex pairs
 on $\mathcal{E}$.
Also
the Tate curve 
$E$ is endowed with a dynamical system of $L$-rational points via its
$p$-adic encoding. In the latter, the  
points of the dynamical system
on $E$
are the $L$-rational points
 given by the $H$-orbits of the ends of $\mathcal{T}_L$
encoding the data of $X_t$.  Hence, for fixed $p$,
the time series $X_t$ has the invariants:
$c=\frac{d}{e}$, $L$, $K$ (cyclotomic),  and $\beta_1=1$.
The corresponding $p$-adic time series obtained by encoding
has the further invariant (assuming that $p^c\in L$)
$$
\Theta=\begin{pmatrix}-p^{c}&0\\1-p^{c}&-1\end{pmatrix}
$$ 
which gives rise to the Tate curve $E=\mathbb{G}_m/\langle\theta\rangle$,
where $\theta\in\PGL_2(L)$ is the M\"obius transformation associated 
to $\Theta$.

\begin{Exa}
Consider the series of dendrograms as symbolically depicted
in Figure \ref{dendroseries}. 
The time series of balances follows the recursion
$$
b(t+1)=b(t)+c_t,\quad 
c_t=\begin{cases}-1,&t\equiv 0\mod 2\\-2,&t\equiv 1\mod 2\end{cases}
$$
In the average, the balance increases each time by $c=-\frac32$. Hence, 
we have a translation on the real line by $c$ with quotient graph
$\mathcal{E}$ a circle, and two vertices on $\mathcal{E}$ representing
the two orbits of the marked vertex $\bullet$ in each dendrogram of Figure
\ref{dendroseries}. The loop obtained is depicted in Figure \ref{loop},
where $v_\Even$ represents the dendrograms at even times $t$,
and $v_\Odd$ at odd $t$.
\begin{figure}
\centering
\setlength{\unitlength}{1mm}
\begin{picture}(20,20)
\put(10,10){\circle{40}}
\put(10,17){\circle*{2}}
\put(3.5,7){\circle*{2}}
\put(7,19.5){$v_\Even$}
\put(-2,3.5){$v_\Odd$}
\end{picture}
\caption{Dynamical system on a loop.} \label{loop}
\end{figure}
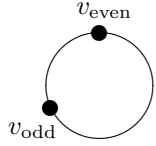
\end{Exa}

\subsection{Mumford curves}

Here, we assume w.l.o.g.\ that $K$ is a sufficiently large $p$-adic number field. By this we mean that the power  $p^u$ by any fraction $u$ which needs to be taken, lies in $K$. This implies that the $u$-th fraction
of a path within $\mathscr{T}_K$ is also defined over $K$, i.e.\
is a sequence of edges from $\mathscr{T}_K$.

\smallskip
Assume that a time series of binary $p$-adic dendrograms $X_t$ gives rise  to
a dynamical system on a Tate curve through an action on the geodesic $]0,1[$
as described in Section \ref{time4Tate}. By translations, we can transform the dendrograms $X_t$
in such a way to $X_t'$ that the segments
$I'_t=T'^\dagger_t\cap]0,1[$
are all balanced, where 
 $T'_t$ is the tree underlying $X_t$. 
If we now assume that $I'_t=[v'_0(t),v'_1(t)]$ 
is approximately constant in time,
then by a small deformation of the family $X'_t$ we may
assume that $v'_0(t)=v'_0=\Const$. Let $e_0(t)$
be the edge originating in $v'_0$ and not lying in $I'_t$.
If further $\mu(e_0(t))$ is approximately constant,
then by another small deformation, we can assume that the time series
$X_t'$ has a fixed vertex $w_0$ which is the target of $e_0(t)$.
This vertex is the root of the time-invariant branch $\Gamma_0'(t)$ of $X'_t$.

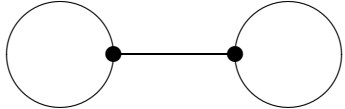
\begin{figure}
\centering
\setlength{\unitlength}{1mm}
\begin{picture}(50,20)
\put(10,10){\circle{40}}
\put(40,10){\circle{40}}
\put(17,10){\circle*{2}}
\put(33,10){\circle*{2}}
\put(17,10){\line(1,0){15}}
\end{picture}
\caption{Segment with two loops.} \label{genus2loop}
\end{figure}

Having made this cascade of assumptions, there is one further assumption
which takes us into the situation of before the introduction of 
time series of genus one. Namely, that $w_0$ lies on a time-invariant geodesic
line $]a,b[$ for some $a,b$ in the data. In fact, the initial terms of the 
two $p$-adic numbers
$a$ and $b$ are uniquely determined by the path $v\leadsto w_0$.
Continuing the $p$-adic expansion with zero coefficients yields $a$,
and continuing with $1$ and then zeros yields $b=a+p^m$ for some
 $m$ larger than the highest power of $p$ occurring in $a$. 
In the case that the
conditions for constructing the Tate curve are fulfilled for the branches
$\Gamma'_0(t)$, we end up  with a $p$-adic riemann surface
of genus $2$ because of the translation $\sigma$ by a fraction $u$ 
along the geodesic 
$]a,b[\subseteq\mathscr{T}_K$.
In fact, $\sigma$ is represented by an hyperbolic transformation
$\varsigma\in\PGL_2(K)$
with matrix
$$
\begin{pmatrix}a-p^ub&(p^u-1)ab\\1-p^u&p^ua-b\end{pmatrix}. 
$$
Together with $\theta$, we obtain a discrete
subgroup $F_2=\langle\theta,\varsigma\rangle$ of $\PGL_2(K)$ generated by
$\theta$ and $\varsigma$. The closure in $\mathbb{P}^1$ of the
union of the $F$-orbits of $0,1,a,b$ is a set $\mathscr{L}$ whose
complement $\Omega=\mathbb{P}^1\setminus\mathscr{L}$ is a $p$-adic manifold on
which $F_2$ acts, and the quotient $C=\Omega/F_2$ is a $p$-adic riemann
surface of genus $2$. A $p$-adic riemann surface of genus $2$ or higher is
usually called a {\em Mumford curve}. The Mumford curve $C$ comes again
 with a dynamical system on its $K$-rational points
$C(K)=\Ends(\mathscr{T}_K/F_2)$ given by the orbits of the data.
The smallest subtree ${T}^*\langle\mathscr{L}\rangle$ of 
$\mathscr{T}_K$ such that
$\Ends(T^*\langle\mathscr{L}\rangle)=\mathscr{L}$ is an
$F_2$-invariant
tree, and the resulting quotient graph $\mathcal{C}$ is a finite graph with 
first Betti number $h_1(\absolute{\mathcal{C}},\mathbb{R})=2$ as illustrated in Figure \ref{genus2loop}.


\begin{Rem}
The fact that the geodesic lines $]0,1[$ and $]a,b[$ in the Bruhat-Tits tree
are disjoint is sufficient for the translations $\tau$, $\sigma$ to generate 
a discrete hyperbolic group
$\langle\theta,\varsigma\rangle\subseteq\PGL_2(K)$, and hence give
rise to a Mumford curve of genus $2$.
This, however, is not a necessary condition. In fact, if the geodesic lines
intersect in a segment $I$, then the length of $I$ must not be larger than
the periods of $\tau$ and $\sigma$ in order for the group of hyperbolic
transformations to be discrete. In the non-discrete case, there is no
Mumford curve obtained by the action on the projective line.
The case of time series with time-invariant intersecting geodesics  
is a bit more involved
and will be treated elsewhere.  
\end{Rem} 
 
\section{Conclusion} 
We have studied dendrograms from the viewpoint of $p$-adic geometry,
where combinatorial objects are associated to spaces in a natural way.
The space here is the $p$-adic projective line $\mathbb{P}^1$,
and punctures of $\mathbb{P}^1$ define a subtree of the $p$-adic
Bruhat-Tits tree, i.e.\ a dendrogram whose data $D$ are the punctures.
This dendrogram is the hierarchic classification of the $p$-adic numbers
from $D$ with respect to the $p$-adic norm $\absolute{\cdot}_p$.
Due to the ultrametric property of $\absolute{\cdot}_p$, the
classification algorithm for $p$-adic numbers is simple. Hence, the focus
in data mining
shifts from classification to $p$-adic data encoding,
a task which in general is
far from trivial. 
However, in the case of strings over a finite alphabet $\mathcal{A}$,
we have observed that the task becomes
much simpler, because the lettres from $\mathcal{A}$
can be identified with coefficients in the $p$-adic expansion of numbers.
Finally, we have introduced the genus $g$ of a time series of $p$-adic
dendrograms by associating to it a discrete 
action on the Bruhat-Tits tree, exemplified in the cases $g=1$ and $g=2$.
From this action, a finite quotient graph $G$ 
can be constructed. Even more, the action
yields a dynamical system
on a so-called {\em Mumford curve} of genus $g$ whose associated
combinatorial object from $p$-adic geometry is $G$. 
These new invariants now await 
 practical application in the study of time series data. 

\section*{Acknowledgement} 
The author acknowledges support from DFG-project BR 3513/1-1
``Dynamische Geb\"aude\-be\-stands\-klas\-si\-fi\-ka\-tion'' and thanks Fionn Murtagh 
for fruitful  discussions  via email and suggestions for improving
the exposition of this article.


\begin{thebibliography}{9} 
\bibitem{BradDegendendrofam}Bradley, P.E. (2006) Degenerating families of dendrograms. \textit{Preprint}.
\bibitem{BradDendrofam}Bradley, P.E. (2007) Families of dendrograms. \textit{Preprint}.
\bibitem{CK2005}Cornelissen, G. and Kato, F. (2005) The $p$-adic icosahedron. \textit{Notices of the AMS}, \textbf{52}(7), 720--727.
\bibitem{DNAp}Dragovich, B. and Dragovich A. (2006) A $p$-adic model of DNA sequence and genetic code. \textit{Preprint},\break arXiv:q-bio.GN/0607018. 
\bibitem{DKM2006}Dragovich, B., Khrennikov, A. and Mihailovi\'c, D. (2006) Linear fractional $p$-adic and adelic dynamical systems. \textit{Preprint}, arXiv:math-ph/0612058.
\bibitem{Gouvea} Gouv\^{e}a, F.Q. (1993) {\em $p$-adic numbers: an introduction}. 
Universitext, Springer-Verlag, Berlin.
\bibitem{KK2007} Khrennikov, A.Y. and Kozyrev, S.V. (2007) Genetic code on the diadic plane. \textit{Preprint}, \break arXiv:q-bio.QM/0701007.
\bibitem{MurtaghJoC2004}Murtagh, F. (2004) On ultrametricity, data coding, and computation. \textit{J.\ of Classification.} \textbf{21}, 167--184.
\bibitem{Murtagh2004} Murtagh, F. (2004) Thinking ultrametrically. In: D.\ Banks, L.\ House, F.R.\ McMorris, P.\ Arabie, and W.\ Gaul (eds.): \textit{Classification, Clustering and Data Mining}, Springer, 3--14.
\bibitem{MDC2006} Murtagh, F., Downs, G. and Contreras, P. (2006) Hierarchical clustering of massive, high dimensional data sets by exploiting ultrametric embedding. \textit{Preprint}.
 \bibitem{Neukirch}Neukirch, J. (1999) {\em Algebraic number theory}. Springer-Verlag, Berlin.
\end{thebibliography}
\end{document}